# Estimating the Attractor Dimension of the Equatorial Weather System[1]

**Melvin Leok Boon Tiong**

*Block 106 Spottiswoode Park Road #05-140 Singapore 080106*
*E-Mail: melvin@singnet.com.sg*
*http://www.singnet.com.sg/~melvin/*

**Abstract**

The correlation dimension and limit capacity serve theoretically as lower and upper bounds, respectively, of the fractal dimension of attractors of dynamic systems. In this paper, we show that estimates of the correlation dimension grow rapidly with increasing noise level in the time-series, while estimates of the limit capacity remain relatively unaffected. It is therefore proposed that the limit capacity be used in studies of noisy data, despite its heavier computational requirements.

An analysis of Singapore wind data with the limit capacity estimate revealed a surprisingly low dimension (~2.5). It is suggested that further studies be made with comprehensive equatorial weather data.

## 1. Introduction

Estimations of the fractal dimension of meteorological time-series serve a fundamental role in the development of dynamic models of meteorological phenomena. These dimensional estimates provide bounds for the number of independent variables necessary to model the system, and assists one in determining the appropriateness of a model.

There have been numerous efforts to determine the dimension of weather and climatic attractors [1,2,3], however the equatorial weather system has yet to be subjected to a thorough analysis. Previous dimensional estimates have been based upon the use of the correlation dimension, but the reliability of this measure is questionable as it is unstable when perturbed by noise in the observational time series.

In this study, the stability of the estimates of the limit capacity and the correlation dimension is investigated by considering the Lorenz attractor perturbed by various levels of noise.

The limit capacity dimension of a 22-year equatorial wind time-series, with a sampling frequency of once a day is then estimated.

## 2. Basic Concepts

The phase portrait of a dynamical system can be reconstructed from the observation of a single variable by the method of delays as proposed by Takens [4]. Takens proved that almost all smooth $d$-dimensional manifolds can be embedded in a $(m=2d+1)$-dimensional space while preserving geometrical invariants. The observational time series $x_1, x_2, x_3, ..., x_N$ is represented by the set of vectors $\mathbf{x}_t = \{x_t, x_{t+\tau}, ..., x_{t+(m-1)\tau}\}$ (where $t=1,2,...,N-[m-1]\tau$) in the reconstructed phase space.

---

The delay $\tau$ is usually chosen for which the autocorrelation function,

$$C(\tau) = \frac{1}{N}\sum_{i=1}^{N-\tau}(x_i - \bar{x})(x_{i+\tau} - \bar{x}) \tag{1}$$

(where $\bar{x}$ is the arithmetic mean) is close to zero, thus minimizing the statistical dependence among the coordinates of the vectors.

In practice, one does not know *a priori* the dimension of the dynamical system, and the embedding dimension which is necessary for the phase space reconstruction. Thus, the dimensional estimate is computed for increasing embedding dimensions until the dimensional estimate stabilizes.

### 2.1. Correlation Dimension

The correlation dimension theoretically provides a lower bound to the fractal dimension of an attractor, thus it satisfies the following inequality

$$d_c \leq d_f \tag{2}$$

The correlation dimension is obtained by considering the cumulative correlation function, defined by Grassberger and Procaccia [5] as

$$C(r) = \lim_{N\to\infty}\frac{1}{N^2}\sum_{\substack{k,j=1\\k\neq j}}^{N}\theta\left(r - \|\mathbf{x}_k - \mathbf{x}_j\|\right) \tag{3}$$

where $\theta$ is the Heaviside function, such that $\theta(x) = 0$ if $x \leq 0$ and $\theta(x) = 1$ if $x > 0$. The Euclidean norm utilized in (3) is defined as

$$\|\mathbf{x}_k - \mathbf{x}_j\| = \sqrt{\sum_{i=1}^{m}(x_{k+(i-1)\tau} - x_{j+(i-1)\tau})^2} \tag{4}$$

The cumulative correlation function is related to the correlation dimension by the power law

$$C(r) \sim r^{d_c} \qquad (r \to 0) \tag{5}$$

This results in the correlation dimension $d_c$ being defined as

$$d_c = \lim_{r\to 0}\frac{\ln[C(r)]}{\ln[r]} \tag{6}$$

### 2.2. Limit Capacity

As opposed to the correlation dimension, the limit capacity, $d_l$ serves as an upper bound to the fractal dimension of an attractor. Considering this property, and the inequality (2), we obtain

$$d_c \leq d_f \leq d_l \tag{7}$$

Let $N(\varepsilon)$ *be the minimum number of spheres of radius $\varepsilon$ necessary to cover all points of the attractor.* Takens [4] defined the limit capacity $d_l$ as follows:

$$N(\varepsilon) \sim \varepsilon^{-d_l} \qquad (\varepsilon \to 0) \tag{8}$$

From the relation (8), we obtain the limit capacity $d_l$ as

$$d_l = \lim_{\varepsilon\to 0}\frac{\ln[N(\varepsilon)]}{\ln[1/\varepsilon]} \tag{9}$$

*2.3. Methods of estimating dimensions*

The reconstruction of an attractor by the method of delays as described in Section 2 is performed on the observational time-series. The embedding dimension *m* which is used in the attractor reconstruction begins from the value 2 and proceeds to successively higher dimensions until the dimensional estimate stabilizes. For the correlation dimension, $C(r)$ vs $r$ is plotted on a log-log graph, and the gradient of the region of the graph which exhibits scaling behaviour yields an estimate of the correlation dimension. Similarly, the limit capacity is obtained by plotting $N(\varepsilon)$ vs $1/\varepsilon$ on a log-log graph, with the gradient providing an estimate of the limit capacity.

### 3. Effect of Noise on Dimensional Estimates

We consider the effect of noise on the correlation dimension and limit capacity estimates by computing the dimensional estimates of the discretely sampled time-series obtained from the *x* component of the Lorenz Attractor, which is described by the following set of ordinary differential equations,

$$\frac{dx}{dt} = \sigma(y-x), \quad \frac{dy}{dt} = rx - y - xz, \quad \frac{dz}{dt} = xy - bz \qquad (10)$$

with the following parameters, $\sigma=10$, $r=28$, $b=8/3$. The Lorenz Attractor is shown in Fig. 1.

The Lorenz equations were numerically integrated with a fourth-order Runge-Kutta method [6], with $\Delta t=0.005$ and $\tau=0.6$ (which was the time lag at which the autocorrelation function for the discrete time-series of the Lorenz attractor vanished). A discrete time-series with 2048 data points of the *x* component was then generated and subjected to dimensional estimation. Each data point of the time-series was perturbed by the introduction of noise with a maximum magnitude which is 1% and 10% of the maximum magnitude of the *x* component. The noise was generated by a pseudo-random multiplicative linear congruential generator proposed by Park and Miller in [7].

As can be seen from Fig. 2, the dimensional estimates of the Lorenz attractor without noise converge to a finite value with 2.06 for the correlation dimension and 2.31 for the limit capacity at an embedding dimension $m=(2n+1)$, thus satisfying inequality (2.2.1). The attractor with 1% noise as shown in Fig. 3, also behaves similarly with a correlation dimension of 2.183 and a limit capacity of 2.342. However, when the time series is perturbed by 10% noise as shown in Fig. 4, inequality (7) is *not* satisfied, as the correlation dimension of 4.678 is *greater than* the limit capacity of 2.332.

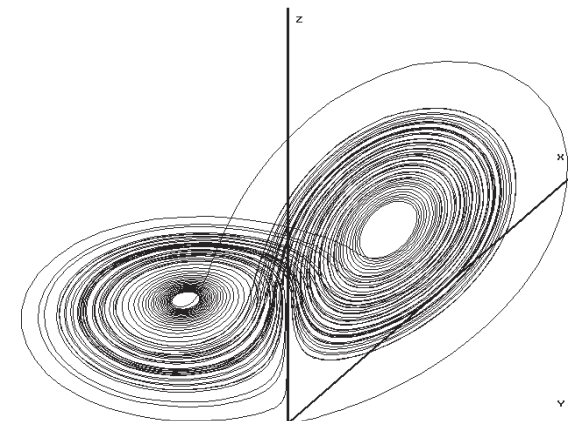
Fig. 1: Lorenz Attractor

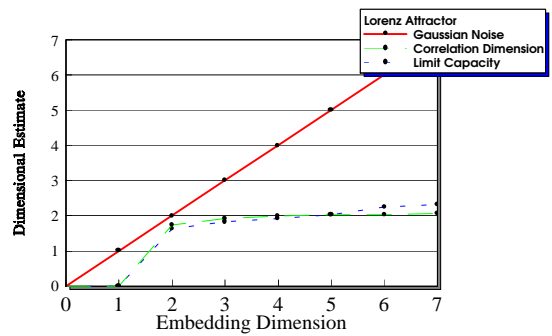
Fig. 2: Dimensional Estimate of Lorenz Attractor

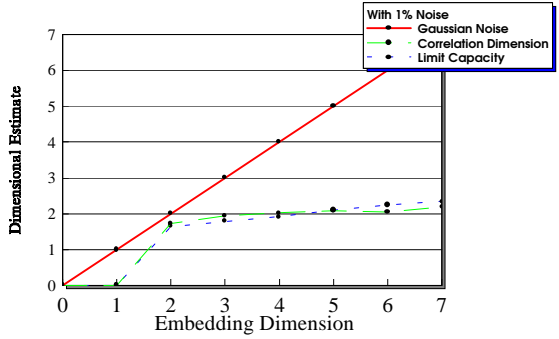
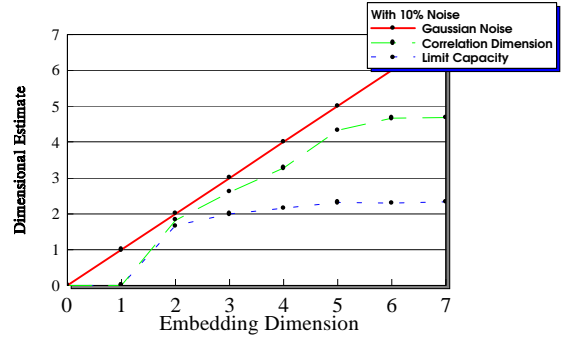

Fig. 3: Dimensional Estimate of Lorenz Attractor with 1% Noise

Fig. 4: Dimensional Estimate of Lorenz Attractor with 10% Noise

*3.1. Analysis of Dimensional Estimates*

The correlation dimension is extremely sensitive to noise, as illustrated by the rate of divergence of the correlation dimension from the dimensional estimate of the noiseless attractor upon the addition of noise. This is significant, since the increase in the correlation dimension due to noise would potentially render the lower bound to the fractal dimension provided by the correlation dimension invalid, as in the case of the Lorenz attractor with 10% noise. Thus it would seem that the correlation dimension would tend to measure the dimension of the noise as opposed to the underlying dynamics which are of interest. A possible explanation for this is that for a noisy attractor, the form of the power law relation (5) is

$$C_m(r) \sim r^m \qquad r < r_{noise}$$
$$C_m(r) \sim r^{d_c} \qquad r > r_{noise} \qquad (11)$$

Given that the scaling region for the correlation dimension occurs in the region ($r \to 0$), it follows that the dimensional estimate would tend to be higher than that inherent in the dynamics of the system.

The analogous situation does not occur in the case of the limit capacity as shown from the small deviation from the dimensional estimate obtained for the noiseless attractor. An intuitive explanation is as follows. In the region of small $r$, the limit capacity estimate suffers from the effect of saturation due to high embedding dimensions and a finite time-series, and hence it is generally necessary to obtain the limit capacity estimate at a moderate $r$ region, where the effect of noise is less dominant.

## 4. Dimensional Estimate of Wind Attractor

The time-series analysed are the $x$ and $y$ components of the daily wind velocity observations over Singapore. Missing data were linearly interpolated. A Fourier transform was performed upon each of the two time-series.

As shown in Fig. 5 and Fig. 6, there is a conspicuous peak in the yearly cycle range. The long term cycles are of such high magnitude that they have to be filtered off to allow for the analysis of the dynamics of the weather for periods of less than 140 days. The autocorrelation function is plotted for this filtered data and is shown in Fig. 7 and Fig. 8. The value of $\tau$ for which the autocorrelation function vanishes is 4 days for the *X-Winds* time-series and 8 days for the *Y-Winds* time-series.

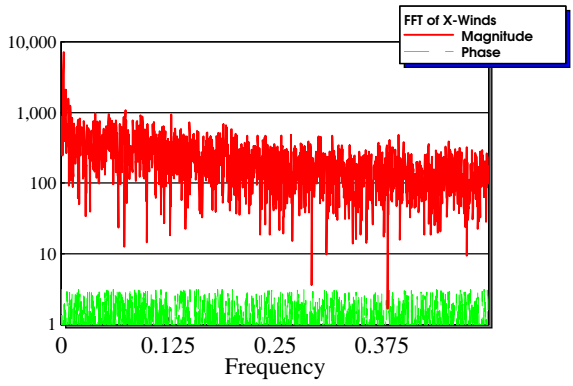

Fig. 5: FFT of X-Winds

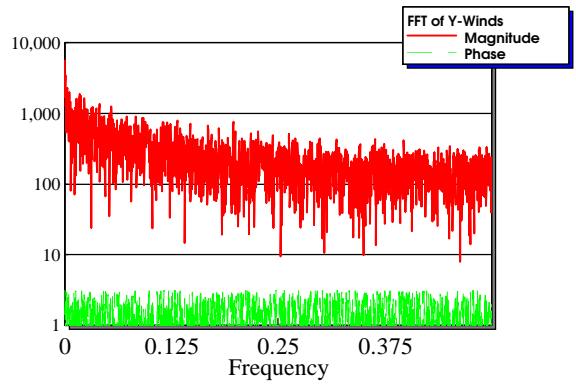

Fig. 6: FFT of Y-Winds

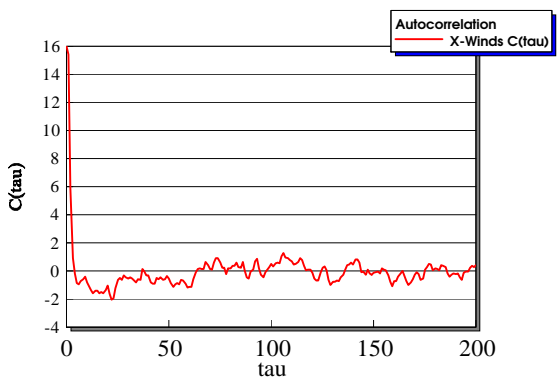

Fig. 7: Autocorrelation of X-Winds

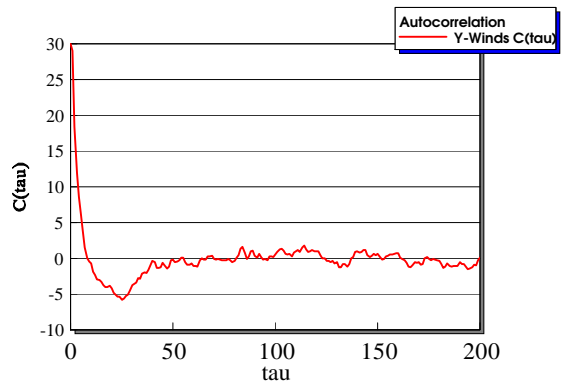

Fig. 8: Autocorrelation of Y-Winds

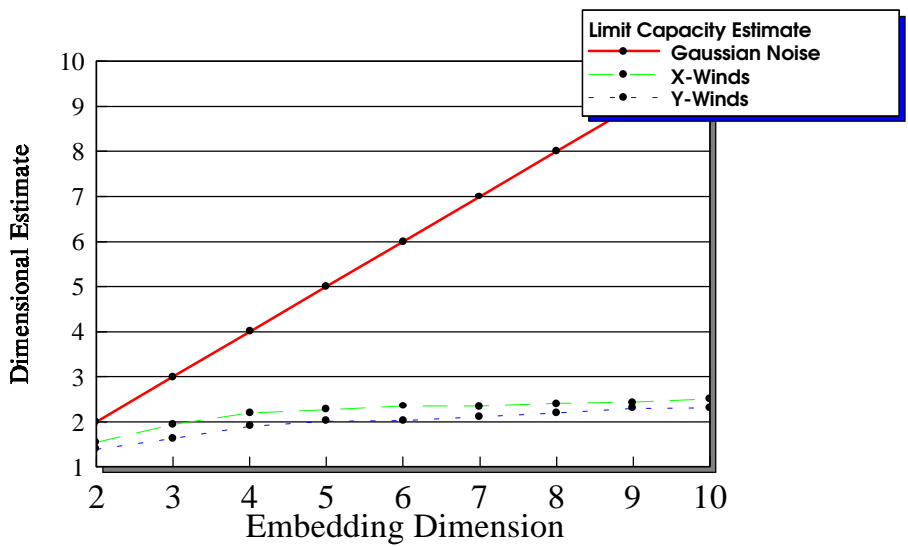

Fig. 9: Limit Capacity Estimate of Wind Data

The limit capacity dimensional estimate was then performed upon the two time-series for embedding dimensions from 2 to 10. The dimensional estimate was obtained by plotting the points $(1/\varepsilon, N(\varepsilon))$ on a log-log graph. The region satisfying the power law relation would be represented by a straight line on the log-log graph. A least-squares fit is then performed to provide the limit capacity dimensional estimate. As seen from Fig. 9, the estimated dimension deviates rapidly from the linear behaviour which would have been expected if the series consisted of Gaussian noise, thus implying that the dynamics are of a deterministic and finite-dimensional nature. It is however essential to realize that the choice of range for the scaling region from which the limit capacity dimensional estimate was derived is somewhat subjective, and a different range could result in a slightly differing dimensional estimate.

| Embedding Dimension | Gaussian Noise | X-Winds | Y-Winds |
|---|---|---|---|
| 2 | 2 | 1.537 | 1.392 |
| 3 | 3 | 1.939 | 1.634 |
| 4 | 4 | 2.190 | 1.901 |
| 5 | 5 | 2.271 | 2.017 |
| 6 | 6 | 2.350 | 2.025 |
| 7 | 7 | 2.343 | 2.109 |
| 8 | 8 | 2.402 | 2.193 |
| 9 | 9 | 2.430 | 2.294 |
| 10 | 10 | 2.508 | 2.313 |

Table 1: Limit Capacity Dimensional Estimate of Wind Data time-series

As seen from Table 1, the limit capacity converges to a fractional dimension between two and three, 2.508 for the *X-Winds* time-series, and 2.313 for the *Y-Winds* time-series. This fractional dimension is indicative of the *sensitive dependence on initial conditions* which characterizes the dynamics of the weather system.

The correlation dimension estimates were also computed, and they exhibited behaviour similar to the dimensional estimates for the Lorenz attractor with 10% noise, and *did not* satisfy inequality (7).

## 5. Conclusion

The dimensional estimates obtained from the correlation dimension should be treated with caution as the dimensional estimates are not robust when the time-series is perturbed by noise. The limit capacity in comparison is less affected by noise and is thus recommended for dimensional estimates of noisy dynamical systems.

The low fractal dimensionality obtained for the equatorial weather attractor suggests that the inherent dynamics could be modelled by a low-dimensional deterministic system. The dimensional estimate for the Singapore wind data is lower than that reported for temperate systems. This is surprising as the equatorial weather systems are usually less organized than the temperate weather systems. A comprehensive study with regional data on a range of weather parameters may provide a deeper insight into the dynamics.

## Acknowledgements

The author wishes to thank the Meteorological Service of Singapore for providing the upper air data used in this study. He would also like to express his gratitude to Mdm Alice Heng Wang Cheng, Mr Chang Ee Chien, Miss Ng Hwee Lang, Miss Koh Siew Lee, and especially to his mentor, Associate Professor Lim Hock, Physics Department, National University of Singapore, for his invaluable help and guidance.